\documentclass[final,5p,times,twocolumn,compress]{elsarticle}

\usepackage{graphicx}% Include figure files
\usepackage[utf8]{inputenc}
\usepackage{subfigure}
\usepackage[T1]{fontenc}
\UseRawInputEncoding
\usepackage{bibentry}
\usepackage{dcolumn}% Align table columns on decimal point
\usepackage{bm}% bold math
\usepackage[mathlines]{lineno}% Enable numbering of text and display math
\graphicspath{{figure/}}
\usepackage{array}
\usepackage{braket}
\usepackage{diagbox}
\usepackage{amsmath,amssymb,amsfonts}%
\usepackage{amsmath}
\usepackage{appendix}
\usepackage{multirow}
\usepackage{bm,color,bbm}

\usepackage{float}
\newcolumntype{.}{D{.}{.}{-1}}
\usepackage{amsopn}
\usepackage[colorlinks,linkcolor=blue, anchorcolor=blue, urlcolor=blue, citecolor=blue]{hyperref}
\usepackage{epstopdf}
\usepackage{natbib}
\setcitestyle{super,comma,open={},close={}}

\usepackage{hyperref}	
\hypersetup{
	colorlinks = true,
	citecolor = {blue},
	linkcolor = {red}
}
\begin{document}

\begin{frontmatter}

\title{High-rate quantum digital signatures network with integrated silicon photonics}

\cortext[cor1]{Corresponding author}
\author[label1,label7]{Yongqiang Du}
\author[label4,label7]{Bing-Hong Li}
\author[label2, label7]{Xin Hua}
\author[label4,label7]{Xiao-Yu Cao}
\author[label1]{Zhengeng Zhao}
\author[label1,label6]{Feng Xie}
\author[label6]{Zhenrong Zhang}
\author[label5,label4]{Hua-Lei Yin\corref{cor1}}
\ead{hlyin@ruc.edu.cn}
\author[label2,label3]{Xi Xiao\corref{cor1}}
\ead{xxiao@wri.com.cn}
\author[label1]{Kejin Wei\corref{cor1}}
\ead{kjwei@gxu.edu.cn}

\address[label1]{Guangxi Key Laboratory for Relativistic Astrophysics, School of Physical Science and Technology, Guangxi University, Nanning 530004, China.}

\address[label4]{National Laboratory of Solid State Microstructures and School of Physics, Collaborative Innovation Center of Advanced Microstructures, Nanjing University, Nanjing 210093, China.} 

\address[label2]{National Information Optoelectronics Innovation Center (NOEIC), Wuhan 430074,  China.}        

\address[label3]{Peng Cheng Laboratory, Shenzhen 518055, China}

\address[label5]{Department of Physics and Beijing Key Laboratory of Opto-electronic Functional Materials and Micro-nano Devices, Key Laboratory of Quantum State Construction and Manipulation (Ministry of Education), Renmin University of China, Beijing 100872, China.}

\address[label6]{Guangxi Key Laboratory of Multimedia Communications and Network Technology, School of Computer, Electronics, and Information, Guangxi University, Nanning 530004, China} 

\fntext[label7]{These authors contributed equally to this paper.}

\large

\begin{abstract}

The development of quantum networks is paramount towards practical and secure communications. Quantum digital signatures (QDS) offer an information-theoretically secure solution for ensuring data integrity, authenticity, and non-repudiation, rapidly growing from proof-of-concept to robust demonstrations. However, previous QDS systems relied on expensive and bulky optical equipment, limiting large-scale deployment and reconfigurable networking construction. Here, we introduce and verify a chip-based QDS network, placing the complicated and expensive measurement devices in the central  relay while each user needs only a low-cost transmitter. We demonstrate the network with a three-node setup using an integrated encoder chip and decoder chip. By developing a 1-decoy-state one-time universal hash-QDS protocol, we achieve a maximum signature rate of 0.0414 times per second for a 1 Mbit file over fiber distances up to 200 km, surpassing all current state-of-the-art QDS experiments. This study validates the feasibility of chip-based QDS, paving the way for large-scale deployment and integration with existing fiber infrastructure.

\end{abstract}

\end{frontmatter}

%\makeatletter
%\def\@cite#1#2{\textsuperscript{{#1\if@tempswa , #2\fi}}}
%\makeatother

%% \linenumbers

%% main text
\section*{\large Introduction}
\vspace{-0.3cm}
\noindent 
Cryptography is widespread in modern society and crucial for numerous applications, including e-commerce, digital currencies, and blockchain, all of which depend on data confidentiality, integrity, authenticity, and non-repudiation. Currently, these applications' security relies heavily on public-key cryptography\cite{1976-Diffie,1978Rivest}, which is believed to be secure against eavesdroppers with limited computational capabilities. However, the security of this cryptographic approach is at risk due to rapid developments in algorithms\cite{1994Shor,2020Boudot-argorithmic} and computational power, particularly in the field of quantum computing\cite{2019Arute,2021Arrazola-QC,2022Philips-QC,2024Bluvstein-quantum-computor}.

\begin{figure*}[]
	\begin{center}
		\begin{tabular}{c}
			\includegraphics[height=8cm]{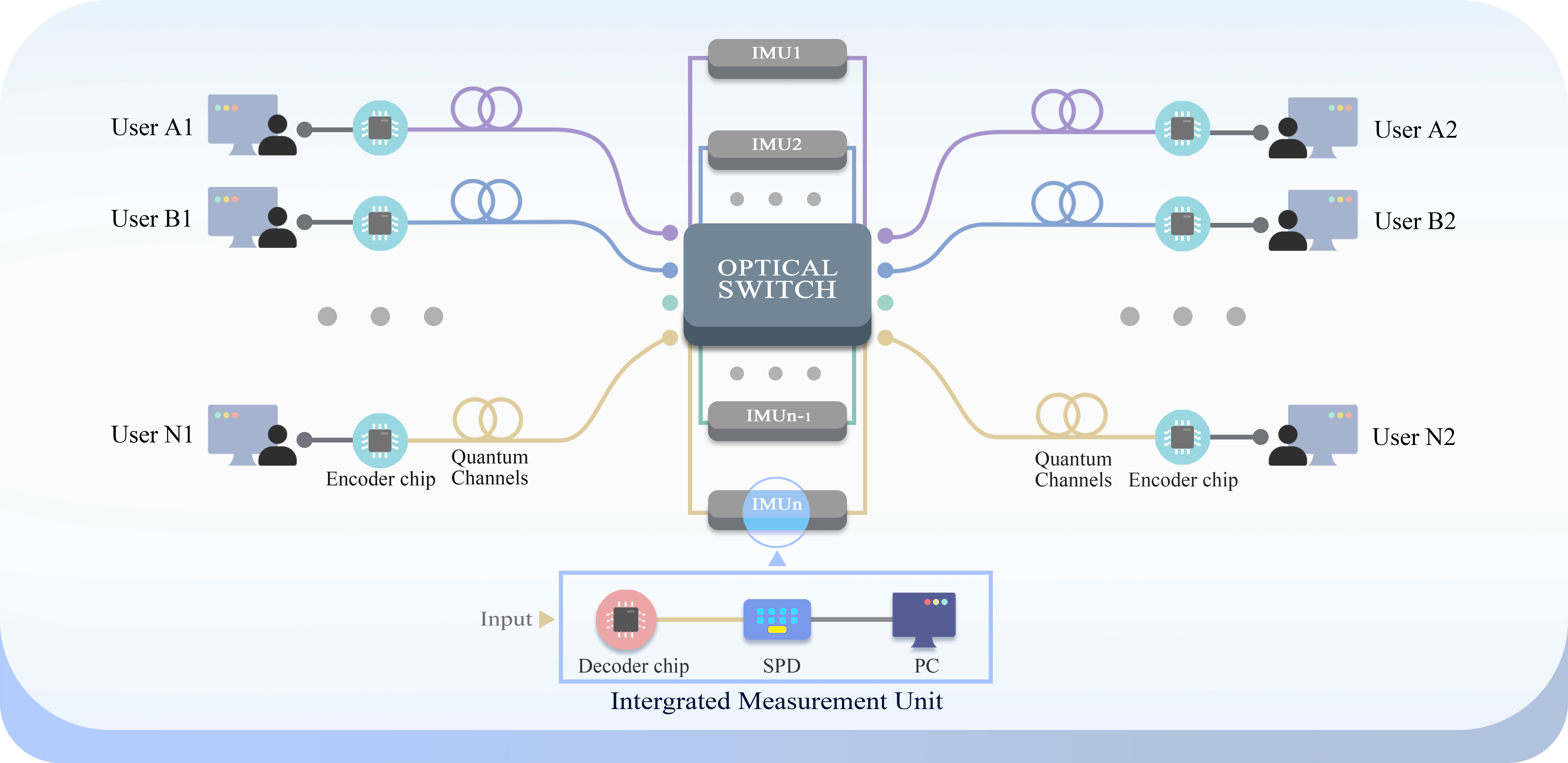}
		\end{tabular}
	\end{center}
	\caption 
	{ \label{Star-network}
		Schematic of the star-like topology chip-based quantum network. It primarily consists of three parts: end-users, optical switches, and integrated measurement units (IMU). Each user holds a transmitting chip to produce quantum states, connected to the optical switch via fiber links.  The optical switch can group users arbitrarily and route the transmitted quantum states to a specified IMU. The IMU mainly comprises quantum state decoding chips, single-photon detectors (SPDs), and personal computers (PC). This network architecture allows for reconfiguring IMU to run BB84 or MDI QDS protocol. } 
\end{figure*}

Unlike classical cryptography, quantum cryptography, utilizing quantum mechanical properties\cite{1984-BB84}, provides a cryptographic toolbox without relying on any assumptions about the computational power of eavesdroppers. A well-known example of quantum cryptography is quantum key distribution (QKD), which offers an information-theoretically secure encryption solution to the key sharing problem, making assumptions only about the devices owned by authorized users\cite{2020-Xu,2020-Pirandola}. With much efforts, QKD has achieved significant milestones, reaching distances of up to 1000 km\cite{2023-liu-TF} and integration into backbone fiber infrastructure of classical communications\cite{2018Mao-Network,2019Dynes-network}.

Different from QKD, quantum digital signature (QDS) enables users to sign documents using quantum methods so that they can be transferred with information-theoretic integrity, authenticity, and non-repudiation. This plays a crucial role in  emails, software distribution, and financial transactions, where data integrity against forgery is paramount. The first QDS protocol was proposed in 2001\cite{2001-Gottesman}, but it was impractical due to the need for long-term quantum storage and secure quantum channels. Substantial efforts have eliminated these impractical technical requirements\cite{2012-Clarke,2014-Dunjko,2016-Yin,2016-Amiri}, and the practical performance of QDS regarding security and signature efficiency has been significantly enhanced\cite{2016-Puthoor,2021-Zhang,2022-Qin}. Specifically, a novel scheme named the one-time universal hash (OTUH)-QDS protocol, first proposed by Yin et al.\cite{2023-Yin} and further developed by Li et al.\cite{2023-Li}, significantly boosts signature efficiency by enabling users to sign messages of any length with information-theoretic security. 

In experiments, QDS has developed from proof-of-principle demonstrations to long distances\cite{2017-Yin-Experimental, 2020-Ding,2021-Richter}, GHz repetition rates\cite{2016-Collins-GHz, 2017-Collins-GHz, 2019-An, 2021-Richter-GHz}, field tests\cite{2016-Croal-field, 2017-Yin-MDI, 2024-Chapman-field}, and reconfigurable networks\cite{2017-Roberts,2022-Pelet,2023-Weng}. These achievements bring QDS closer to maturity and are believed to be the next step in commercial quantum technologies~\cite{2024-Cao}. However, all previous works rely entirely on bulky and expensive optical setups, encountering significant challenges for wide deployments and easy integration of QDS with existing  backbone fiber infrastructures.

In this work, we introduce and verify a chip-based QDS network. In such a network, each user requires only an integrated photonic transmitter chip, while the complex and expensive measurement devices are placed in the central node. We further address the technical challenges of building such a network by developing the 1-decoy-state OTUH-QDS protocol, which allows efficient signatures using one decoy state and non-privacy-amplification keys. This dramatically reduces the manufacturing complexity of the transmitter chip and reduces the computational cost and latency of the post-processing stage. We demonstrate the network with a three-node setup that achieves a maximum signature rate of 0.0414 times per second (tps) for 1 Mbit messages over fiber distances  up to 200 km. This signature rate surpasses current state-of-the-art QDS experiments. This study validates the feasibility of chip-based QDS, paving the way for large-scale deployment and integration with existing fiber infrastructures.

\section*{\large Results}\label{Results}
\vspace{-0.3cm}
\noindent 
\textbf{Network structure}

\noindent 
The schematic of our proposed QDS network is shown in Fig.~\ref{Star-network}. The network features a star-like topology and consists of three main components: end-users, optical switches, and integrated measurement units (IMUs). Each user at the terminal nodes of the network has a compact transmitter chip and is linked to a central node containing several IMUs, including quantum state decoding chips, single-photon detectors (SPDs), and personal computers (PCs).
To implement the QDS task, which typically involves three parties (namely a signer, a verifier, and a receiver), optical switches or dense wavelength division multiplexers are used to arbitrarily group two users as a verifier and a receiver and route the transmitted quantum states to a specified IMU, which is owned by a signer.

\begin{figure*}[t]
	\begin{center}
		\begin{tabular}{c}
			\includegraphics[height=8cm]{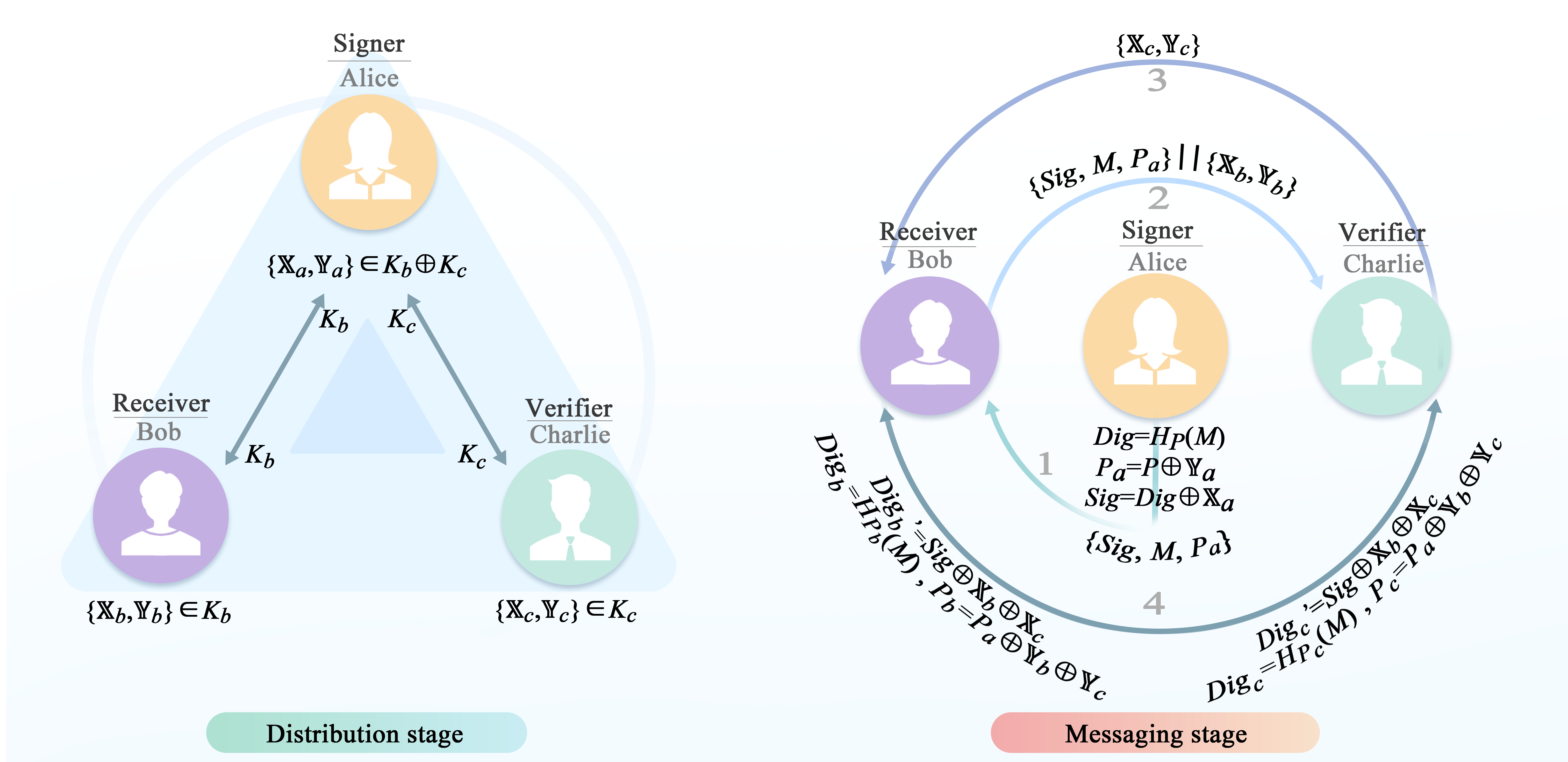}
		\end{tabular}
	\end{center}
	\caption {Schematic diagram for the efficient QDS Protocol. This protocol can be divided into two stages, known as distribution state and messaging state.}
	\label{Protocol}
\end{figure*}

This architecture offers three main advantages. Firstly, each user only needs a compact transmitter chip fabricated by integrated platforms. Integrated photonics provides highly robust manufacturing processes that help reduce costs for personal devices and enable the miniaturization of components and circuits for handheld and field-deployable devices. Secondly, the signer holds the expensive and bulky measurement system, which is shared by all terminal node users, thus bypassing the challenging technique of integrating single-photon detectors on a chip\cite{2021-Gyger,2021-Lomonte}, as users do not need to perform quantum detection. Thirdly, this network architecture easily integrates into existing  classical telecommunications infrastructure and is compatible with the current quantum network by flexibly configuring the IMU\cite{2013-Fröhlich, 2021-Chen-satellite, 2022-Fan-Yuan, 2024-Huang}. For example, the network can upgrade to a measurement-device-independent version by using a Bell-state measurement device~\cite{2017-Roberts,2020-Wei,2020Cao-MDI,2020-Semenenko}.

\vspace{0.5cm}
\noindent 
\textbf{Efficient QDS Protocol}

\noindent 
In order to enhance the performance and compatibility with chip-based network structures, we develop a modified QDS protocol described in Ref.~\cite{2023-Yin,2023-Li} which is capable of signing an arbitrarily long document with imperfect pre-distribution keys.  Our main modification is using 1-decoy-state method in the pre-distribution state. This modification is crucial for building a practical QDS network:

 Firstly, as this protocol does not require the preparation of a vacuum state, it lowers the extinction ratio requirement for IM on a silicon-based chip, thereby decreasing the manufacturing complexity of the transmitter chip. Secondly, as this protocol requires fewer resources for the generation and management of decoy states, it further reduces system complexity. Thirdly, due to its ability to directly utilize imperfect keys without privacy amplification, our protocol can significantly reduce the computational costs and latency associated with post-processing.

Here, we summarize the main steps of our protocol. Details of the specific implementation and the security proof can be found in \ref{QDS} and \ref{sec_ana}, respectively. Our protocol, depicted in Fig.~\ref{Protocol}, is divided into two stages: distribution and messaging stage, outlined as follows.

\textit{Distribution stage:} Bob (Charlie) and Alice share a sequence of raw keys by implementing the 1-decoy-state BB84 key generation protocol (KGP) to create a key bit string $K_b$ ($K_c$) of length $n_{Z}$. That is, Bob (Charlie) sends a signal state with average intensity $\mu$ and probability $P_\mu$, and a decoy state with average intensity $\nu$ and  probability $P_\nu$. Alice then measures them using the basis $\{Z, X\}$ and generates $K_b$ ($K_c$) with an error correction algorithm.

Alice subsequently creates a key string $K_a$ by performing an XOR operation on the key strings $K_b$ and $K_c$. To sign a message $M$, Alice randomly selects $2L$ bits from $K_a$ to create two $L$-bit key strings $\left \{ \mathbb{X_\mathit{a}}, \mathbb{Y_\mathit{a}} \right \}$ and shares the positions of these bits with Bob (Charlie) via an authenticated channel.   $L$ is decided by preset system security parameters $\{\epsilon_{\rm{rob}}, \epsilon_{\rm{rep}}, \epsilon_{\rm{for}}\}$.  Bob and Charlie then independently generate key strings $\left \{ \mathbb{X_\mathit{b}}, \mathbb{Y_\mathit{b}} \right \}$ and $\left \{ \mathbb{X_\mathit{c}}, \mathbb{Y_\mathit{c}} \right \}$, respectively. The three parties then initiate the messaging stage.

\textit{Messaging stage:} Alice creates an $L$-bit digest $Dig$ for the message $M$ using a generalized division hash operation $Dig=H_{P}(M)$, characterized by a local random sequence \(P\). She then encrypts $Dig$ and $P$ using key strings $\mathbb{X_\mathit{a}}$ and $\mathbb{Y_\mathit{a}}$, obtaining $P_a=P \oplus \mathbb{Y_\mathit{a}}$  and the signature $Sig=Dig \oplus \mathbb{X_\mathit{a}}$. Alice transmits $\{Sig, M, P_a\}$ to Bob via an authenticated channel. 

Upon receiving the string, Bob forwards $\{ Sig, M, P_a\}$ along with $\left \{ \mathbb{X_\mathit{b}}, \mathbb{Y_\mathit{b}} \right \}$ to Charlie. Subsequently, Charlie also transfers $\left \{ \mathbb{X_\mathit{c}}, \mathbb{Y_\mathit{c}} \right \}$ to Bob. Bob (Charlie) then independently generates an expected digest $Dig_{b}^{'}=Sig\oplus \mathbb{X_\mathit{b}} \oplus \mathbb{X_\mathit{c}}$ ($Dig_{c}^{'}=Sig\oplus \mathbb{X_\mathit{b}} \oplus \mathbb{X_\mathit{c}}$) and an actual digest $Dig_{b}=H_{P_b}(M)$ ($Dig_{c}=H_{P_c}(M)$), where $P_b=P_a \oplus \mathbb{Y_\mathit{b}} \oplus \mathbb{Y_\mathit{c}}$ ($P_c=P_a \oplus \mathbb{Y_\mathit{b}} \oplus \mathbb{Y_\mathit{c}}$). They then verify the digests. If $Dig_{b}^{'}=Dig_{b}$ and $Dig_{c}^{'}=Dig_{c}$, the signature is accepted; otherwise, it is rejected.

Considering the security framework of a one-time universal hash\cite{2023-Yin,2023-Li} and the 1-decoy-state BB84 KGP\cite{2018-Rusca}, we can define the achievable signature rate as
\begin{equation}\label{eq_SR} R_S=\frac{n_{Z}}{2Lt}, \end{equation}
Here, $t$ represents the cumulative time required to obtain a raw key of length $n_{Z}$.

\begin{figure*}[]
	\begin{center}
		\begin{tabular}{c}
			\includegraphics[height=6cm]{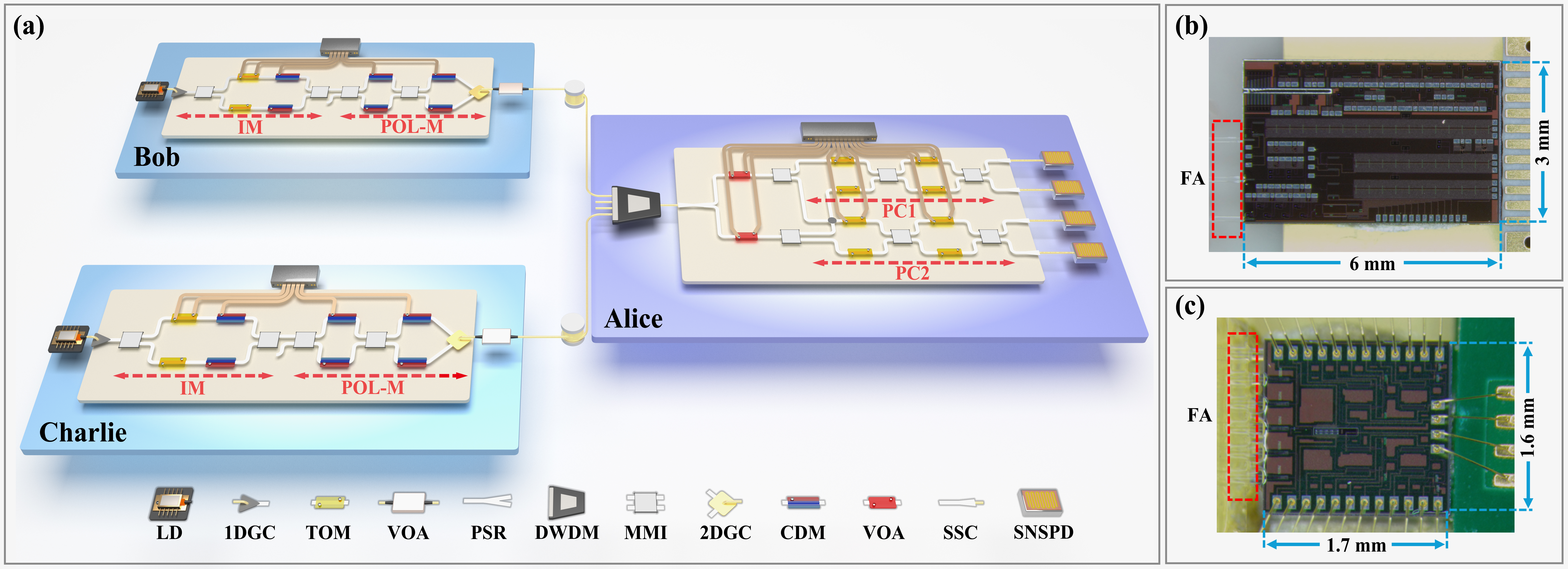}
		\end{tabular}
	\end{center}
	\caption 
	{ \label{Chip-QDS}
		(a) Experimental setup of chip-based three-node quantum network. It mainly consists of two quantum links: Bob-Alice and Charlie-Alice, connected by spooled fibers. The signal lights from Bob and Charlie are multiplexed by a dense wavelength division multiplexer (DWDM) before being measured by Alice. Bob and Charlie each have a laser diode (LD), a silicon-based encoder chip, and a variable optical attenuator (VOA). Alice has a DWDM, a silicon-based decoder chip with polarization tracking capability, and four single-photon detectors (SNSPDs). The encoder chip includes a one-dimensional grating coupler (1DGC) and a two-dimensional grating coupler (2DGC) for optical pulse input and output. The intensity modulator (IM) and polarization modulator (Pol-M) consist of  multimode interferometers (MMIs),  thermo-optical modulator (TOMs) and carrier-depletion modulator (CDMs). The decoder chip includes a mode-size converter (SSC) for optical pulse input and output, a polarization splitter-rotator (PSR) for converting polarization information into on-chip path information, VOA for balancing polarization-dependent loss, MMI and TOM for polarization controller (PC). (b) Microscopic view of the encoder chip.  (c) Microscopic view of the decoder chip. The encoder and decoder chip are coupled to fiber array (FA). } 
\end{figure*}

\vspace{0.5cm}
\noindent 
\textbf{Experimental Setup}

\noindent 
To validate the network, we construct a three-node quantum network, as shown in Fig.~\ref{Chip-QDS}. It includes a central node, Alice, acting as the signer, and two sub-nodes, Bob as the receiver and Charlie as the verifier.

  Bob and Charlie, each using a laser diode (LD), generate phase-randomized light pulses with a repetition rate of 50 MHz and a pulse width of 200 ps. The center wavelengths are 1551 nm and 1549 nm, respectively. Each user's light pulses are coupled into a homemade silicon-based polarization encoder chip. A one-dimensional grating coupler (1DGC) is used for the fiber-to-chip connection. The first structure in the chip, an intensity modulator (IM), generates signal or decoy states. It is implemented via a Mach-Zehnder interferometer (MZI), comprising two multimode interferometers (MMIs), a pair of thermal optical modulators (TOMs) providing static phase bias, and a pair of carrier depletion modulators (CDMs) for dynamic modulation.
  
  The output of the IM is connected to a polarization modulator (Pol-M) used for polarization modulation. The Pol-M comprises an inner MZI driven by a pair of CDMs, bridging an external pair of CDMs, and finally connecting to a two-dimensional grating couplers (2DGC). The 2DGC converts the path-encoding information, modulated by the two pairs of CDMs, into polarization-encoding information and couples it into the fiber. The Pol-M can prepare the four BB84 polarization states, $\left| \psi \right\rangle = ( \left| H \right\rangle + {e^{i\theta }}\left| V \right\rangle)/\sqrt{2}$, where $\theta \in \{0, \pi/2, \pi, 3\pi/2\}$, with $\theta \in \{0, \pi\}$ representing the $Z$ basis and $\theta \in \{\pi/2, 3\pi/2\}$ representing the $X$ basis. The modulated polarization-encoded pulses are attenuated to the single-photon level by a variable optical attenuator (VOA) and then sent to Alice via the fiber channel.

Alice multiplexes the signal photons received from different nodes and couples them into her decoder chip using dense wavelength division multiplexing (DWDM) and an on-chip spot-size converter (SSC). The polarization information carried by the photons is subsequently converted into on-chip path information using a polarization splitter-rotator (PSR). The signal photons are then passively directed for measurement in the $Z$ or $X$ basis using two symmetric MMIs. The measurement of the photons in the $Z$ and $X$ bases is implemented by polarization controllers PC1 and PC2. Each polarization controller (PC) consists of a pair of thermal optical modulators (TOMs) and a Mach-Zehnder interferometer (MZI) driven by another pair of TOMs. By carefully adjusting the drive voltage of the TOMs with a programmable linear DC source, PC1 and PC2 can perform measurements of the quantum state in the $Z$ and $X$ bases. 

The photons measured by the polarization decoder chip are coupled to the external fiber through an SSC and detected by four commercial superconducting nanowire single-photon detectors (SNSPDs). These detectors have a detection efficiency of $70\%$, a dark count rate of approximately 30 Hz, and a dead time of 100 ns. The detection results are registered using a high-speed time-to-digital converter (TDC) and post-processed using a personal computer.

The encoder and decoder chips are realized using standard building blocks provided by a commercial fabrication foundry and are ready for large-scale production. All chips are packaged to protect them from the external environment and enable long-term operation. The size of the encoder chip is 6 $\times$ 3 mm$^{2}$, as shown in Fig.~\ref{Chip-QDS} (b), and it is butterfly packaged with a volume of 20 $\times$ 11 $\times$ 5 mm$^{3}$. The size of the decoder chip is 1.6 $\times$ 1.7 mm$^{2}$, as shown in Fig.~\ref{Chip-QDS} (c), and it is packaged using a chip-in-board assembly with a size of 3.95 $\times$ 2.19 $\times$ 0.90 cm$^{3}$.

\begin{table*}[h!]
	
	\caption{Comparison of recent QDS experiments. $L$ represents the signature length, $\epsilon$ is the security parameter, and $R_S$ represents
		the signature rate.}
	\centering
	\resizebox{\linewidth}{!}{
		\fontsize{10}{10} \selectfont
		\setlength\tabcolsep{12pt}
		\renewcommand\arraystretch{1.5}
		\begin{tabular}{ccccccccc} 
			
			\hline\hline
			References&  \rule[-1ex]{0pt}{3.5ex}Protocol & Clock rate & Distance & Document size& Chip &$L$ &$\epsilon$ & $R_S$  
			\\ \hline
			Roberts et al.~\cite{2017-Roberts}& MDI &1 GHz  &50 km & 1 bit & No & $2.11\times 10^{6}$&$10^{-10}$ & $2.22\times 10^{-2}$ tps
			\\ \hline
			An et al.~\cite{2019-An}& BB84 &1 GHz & 125 km & 1 bit&  No&436490 & $10^{-10}$ & $4.41\times 10^{-2}$ tps
			\\ \hline
			Ding et al.~\cite{2020-Ding}& BB84 &50 MHz & 204 km & 1 bit & No & $4.14\times 10^{10}$& $10^{-5}$ &0.01 tps
			\\ \hline
			Richter et al.~\cite{2021-Richter}& CV &1 GHz & 20 km & 1 bit & No&$2.08\times 10^{8}$ &$10^{-4}$ & 5 tps
			\\ \hline
			Yin et al.~\cite{2023-Yin}&BB84 &200 MHz & 101 km & $ 10^6$ bits & No &256 & $10^{-32}$ &0.82 tps
			\\ \hline
			\multirow{2}{*}{Our work}&\multirow{2}{*}{BB84} &\multirow{2}{*}{50 MHz} &  100 km & \multirow{2}{*}{$ 10^6$ bits} & \multirow{2}{*}{Yes} & 783&$4.64\times 10^{-8}$ & 6.50 tps \\
			& & &  200 km &  &  & 1029& $4.72\times 10^{-8}$ &$4.14\times 10^{-2}$ tps
			
			\\ \hline\hline
			
		\end{tabular}
	}
	\label{table_experi_compare}
\end{table*}

\vspace{0.5cm}
\noindent 
\textbf{QDS for different fiber lengths.}

\noindent 
Using the described setup, we perform a series of QDS experiments and use the example of signing a 1 Mbit messages to demonstrate the performance of QDS.  For each distance, we conduct a numerical optimization to obtain the implementation parameters to enhance the performance of key extraction between each node in the network. For example, at a distance of 150 km, Bob's (Charlie's) intensities of the signal and decoy states are  $\mu=0.597~(0.478)$ and $\nu=0.146~(0.127)$, respectively. The probabilities of sending signal state  $ \mu $ and sending decoy state  $ \nu $ are set to  $P_\mu=0.808~(0.773)$ are $P_\nu=0.192~(0.227)$, and the probability of choosing the state in $Z (X)$ is $P_Z=0.947~(0.934)$ and $P_X=0.053~(0.066)$ , respectively.

\begin{figure}[t]
	\begin{center}
		\begin{tabular}{c}
			\includegraphics[height=6.5cm]{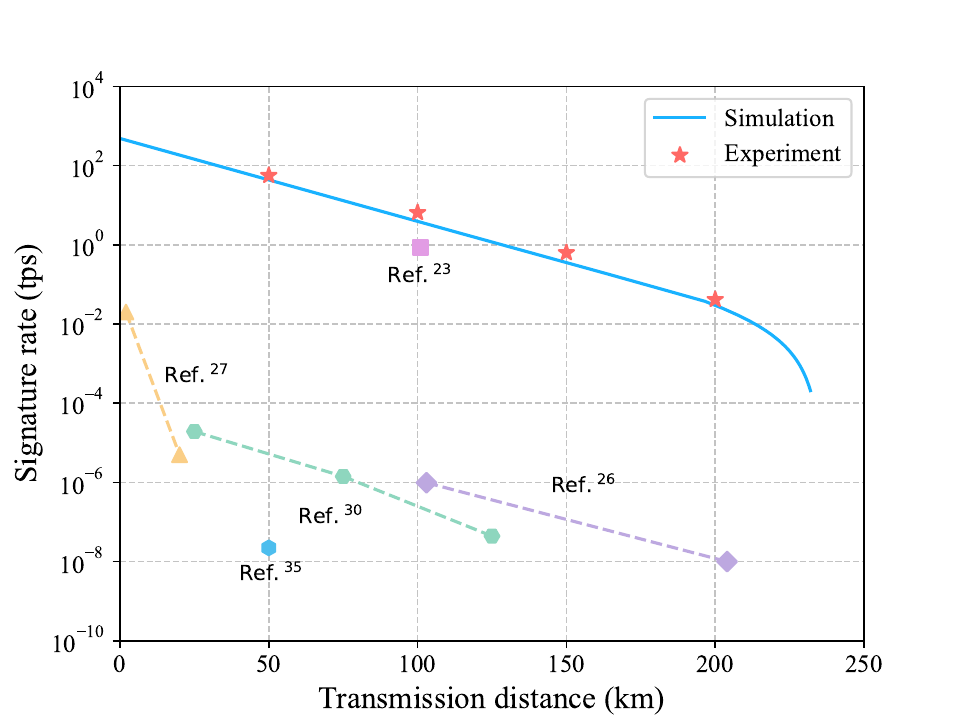}
		\end{tabular}
	\end{center}
	\caption {Simulation and experimental signature rates under different transmission losses. The blue solid line represents the simulated results based on the experimental parameters, while the red dots represent the experimental results at 50 km, 100 km, 150 km, and 200 km  with $n_Z=10^7$  for finite key analysis. We also plot the highest signature rates of current QDS experiments for comparison. Note that the signature rates of our work and Ref.~\cite{2023-Yin} are for signing about 1 megabit files, while the others are for signing 1 bit, where signing multi-bit files can only be achieved by simply repeating the process.}
	\label{SR}
\end{figure}

The experimental results are plotted in Fig.~\ref{SR} and detailed experimental data are provided in \Ref{Exp-R}. %Here, we set the system security parameter $\epsilon =   5\times 10^{-8}$, i.e., $ max\{\epsilon_{\rm{rob}},\epsilon_{\rm{rep}},\epsilon_{\rm{for}}\} \le 5\times 10^{-8}$ and $n_Z$ is set to $10^7$ for finite key analysis.
 It can be seen that we enable to perform secure signature over different fiber spools.  Particularly, we just need to use  2058-bit  key to sign documents with a security bound of $4.72\times10^{-8}$ and a signature rate up to 0.0414 tps over a fiber length of 200 km.

To demonstrate the progress entailed by our results, we compare our experimental results with current state-of-the-art QDS experiments, as shown in Fig.~\ref{SR}. See Table~\ref{table_experi_compare} for a detailed comparison. Our experiment reports the highest signature rate for QDS using the first chip-based setup. Additionally, our revised protocol achieves a higher signature rate than that reported in Ref.~\cite{2023-Yin}, despite our setup having a lower repetition rate.

\begin{figure*}[h]
	\begin{center}
		\begin{tabular}{c}
			\includegraphics[height=9.5cm]{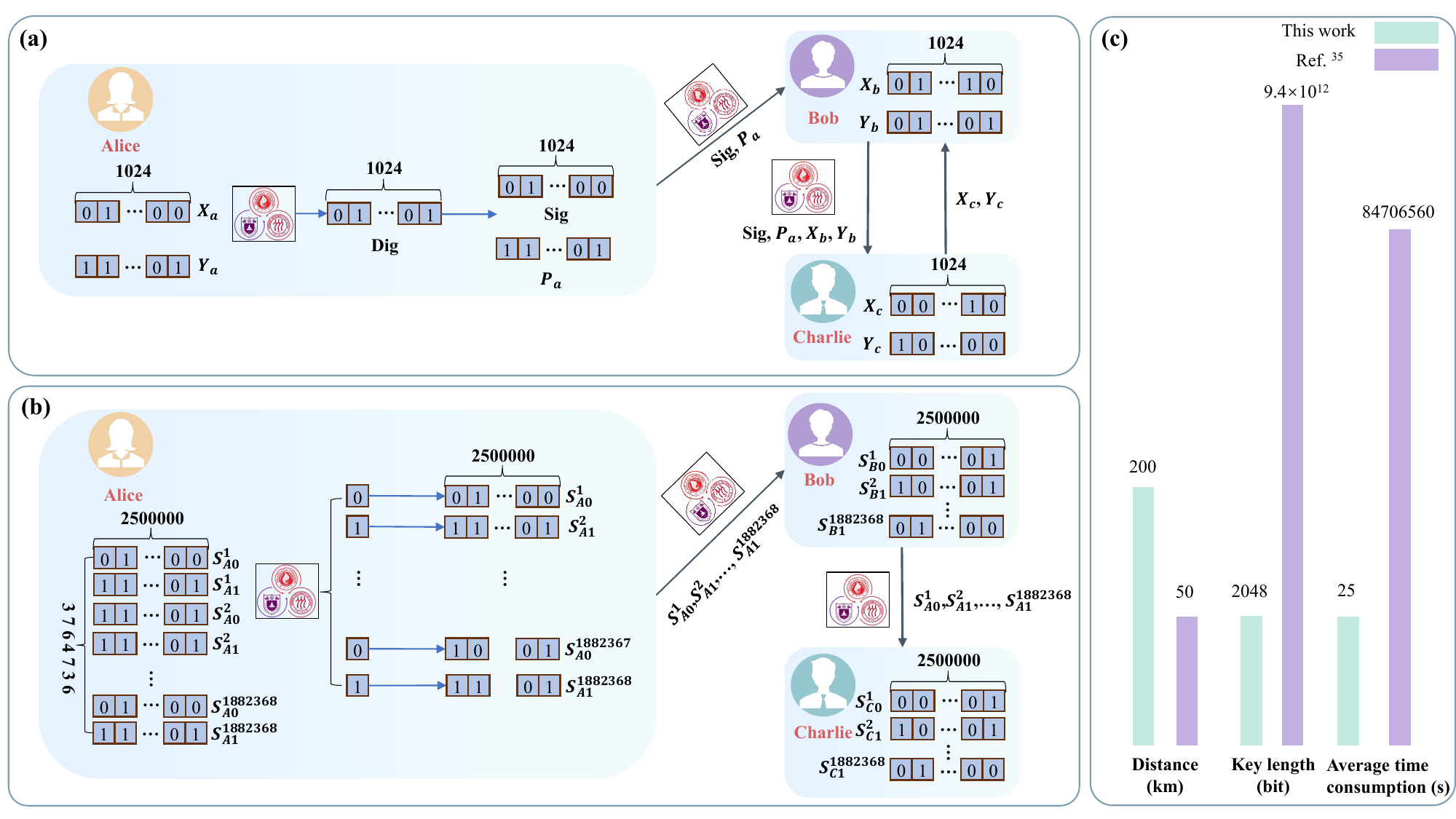}
		\end{tabular}
	\end{center}
	\caption 
	{ \label{illustration}
		(a) An illustration of OTUH-QDS based on the experimental results. A picture including the school emblems of Nanjing University, Guangxi University, and Renmin University of China with a size of 1882368 bits ($258\times 304\times 3\times 8$ red green blue (RGB) color code) is signed. (b) The procedure of signing the same document as in (a) based on scheme in ref.~\cite{2017-Roberts}. (c) Comparison of two schemes on distance, signature length and average time consumption. The green represents this work and purple represents that in ref.~\cite{2017-Roberts}.} 
\end{figure*}

To further illustrate our results, we compare the proposed scheme with the current state-of-the-art QDS system~\cite{2017-Roberts} on a digital signature task for a file of approximately 1.88 M in size. The visual illustration is shown in Fig.~\ref{illustration}. Our work exhibits a simple signature process capable of signing arbitrarily long documents. In contrast, the work reported in Ref.\cite{2017-Roberts} requires performing a one-bit-one signature process. Furthermore, even over longer distances (200 km vs. 50 km), our work requires only 2048 bits with an average accumulation time of 25 seconds to sign documents, while the work reported in Ref.\cite{2017-Roberts} requires $9.4 \times 10^{12}$ bits with an average accumulation time of $8.4 \times 10^{7}$ seconds.

\noindent

%\section*{\large Conclusion}\label{Conclusions}
%\vspace{-0.3cm}
%\noindent 

%In this work, we demonstrated a chip-based three-node QDS network. Each user requires only a low-cost transmitter chip, with the measurement device centralized at the central node, bypassing the technical challenge of integrating SPD into the chip.  The experimental results show that our scheme significantly outperforms previous systems in terms of signature rate. Our results pave the way for a low-cost, wafer-scale manufactured QDS system and offer a promising approach for integrating QDS into future quantum networks.

% \section{DEVICE DESIGN AND FABRICATION}\label{appendix-chip-design}
 \section*{\large Methods}\label{METHODS}
\vspace{-0.3cm}
\noindent 
\textbf{Characterization of components} 

\noindent 
At a repetition rate of 50 MHz, the IM in the encoder chip held by Bob (Charlie) achieved a static extinction ratio of approximately 27 dB (25 dB) through driving TOM and a dynamic extinction ratio of about 18 dB (19 dB) through driving CDM. These parameters meet the requirements of the 1-decoy-state scheme\cite{2018-Rusca}. the polarization states generated by the polarization extinction ratio of Alice's (Bob's) Pol-M is approximately 23 dB (19 dB). The performance of encoder chip  ensures the implementation of a low-error-rate and highly stable key generation.

To characterize the performance of the encoder chip under high-speed modulation, we measure the 3 dB bandwidth of CDM by observing eye diagrams, and the highest value reached approximately 10 GHz. This indicates our setup can support high-speed quantum state preparation with upgrading electronics control. The 3 dB bandwidth of TOM on both the encoder and decoder chips is around 3 kHz. This enables the decoder chip to provide rapid polarization tracking in field-buried and aerial fiber channel scenarios.

The measurement terminal in the Bob (Charlie)-Alice link mainly consists of a DWDM and a decoder chip, causing a total loss of approximately 8.3 dB (8.9 dB). This includes a 0.9 dB (1.5 dB) loss from the DWDM and approximately 7.4 dB loss from the decoder chip. These losses reduce the efficiency of detecting photons, thereby affecting the bit rate of the key generation system.

\vspace{0.3cm} 
\noindent 
\textbf{Generalized division hash functions}

\noindent 
In our protocol, we use generalized division hash functions to   divide the input document.  A generalized division hash function used in the protocol is decided by an irreducible polynomial of order $L/8$ in  Galois Field (GF) (256). Note that $L/8 \times \log_{2}{256}=L$, and the parameter 256 can also be set as other number. If the polynomial is $P(x)$ and the input document is $M$, the hash function is defined as $h(M)= M(x)x^{L/8}$ mod $P(x)$, where all calculations are on GF(256) and $M(x)$ is the polynomial generated by transforming every eight bits of $M$ into its coefficient in turn. Note that every eight bits can be naturally mapped into an element in GF(256) through an isomorphism. The output is also a polynomial with order no more than $L/8-1$, and thus can be characterised into an $L/8$-element GF(256) array, and transformed into an $L$-bit string. In the demonstration we choose the $L$-bit string as the final output of the hash function, i.e., $Dig$. 

\vspace{0.3cm}
\noindent 
 \textbf{Error correction algorithms}

\noindent 
 We conducted an error correction process to obtain the identical raw keys after the key generation process. The detailed process is presented as follows:

(1) The length of the key for each round of error correction is set to $10^6$. The two parties in need of error correction use prepared random sequences to shuffle the original key sequence, and set the length of each segment to $0.73/E_Z$ based on the bit error rate $E_Z$~\cite{2022-mao}. 

(2) Each party calculates the values of parity check node of each segment and compares them with each other through a publicly authenticated channel. When differences are found, binary search is used to locate the error position. The segment with different value is divided into two equal-length blocks and the parity check codes of the two blocks are disclosed. For blocks with different parity check code values, binary search continues until the error position is located, and then one of the participants flips this bit. When a parity check code value is disclosed, the amount of information leakage increases by one.

(3) In the second round, random shuffling and segmentation continue. The block length can be updated to reduce information leakage according to the new error rate. Similar to the first round, the parity check codes of each block are compared, and errors are located using binary search. Additionally, based on newly discovered error positions, errors in the first round blocks are located. 

(4) The above steps are repeated multiple times for the second round. Typically, two additional rounds are enough. 

The error correction efficiency is always lower than 1.16 during our implementation while we set it 1.16 during simulation.

\appendix    % this command starts appendixes

\section{1-decoy-state OTUH-QDS protocol}\label{QDS}

Here, we provide a detailed explanation of the specific steps involved in the distribution stage and the messaging stage of the 1-decoy-state OTUH-QDS protocol.

\vspace{0.3cm} 
\noindent 
\textbf{Distribution stage.}  Bob and Charlie will independently implement the 1-decoy-state BB84 key generation protocol (KGP) with Alice to distribution a pair of identical raw bit strings. The specifics are as follows:

\textit{1. Transmission.} Bob (Charlie) randomly selects a bit value, chooses an intensity choice $k \in \left \{ \mu, \nu \right \}$ (\{ signal, decoy \} ) with probabilities $P_k \in \left \{ P_\mu, P_\nu \right \}$, and selects a basis $\lambda \in \left \{ Z, X \right \}$  with probabilities $P_\lambda \in \left \{ P_Z, P_X \right \}$. Based on these selections, Bob (Charlie) prepares a phase-randomized weak coherent pulse from four states $\left \{ \left | 0_Z  \right \rangle, \left | 1_Z  \right \rangle, \left | 0_X  \right \rangle, \left | 1_X  \right \rangle  \right \} $, where $\left \{ \left | i_\lambda   \right \rangle  \right \}$  represents the state corresponding to bit "$i$" in the "$\lambda$" basis. Finally, Bob (Charlie) sends this prepared state to Alice via the quantum channel.

\textit{2. Detection.} Alice selects a basis from $\left \{ Z, X \right \} $ with probabilities $\left \{ P_Z, P_X \right \}$ to measure the received pulses. For events resulting in a detection click, she records both the basis choice and the corresponding measured bit value. (For double-click events, she assigns a bit value at random.)

\textit{3. Basis reconciliation.} Bob (Charlie) and Alice  publicly announce their basis and intensity choices via an authenticated channel. Then, they determine the number of detection events  $n_{\lambda, k}$, when both Bob (Charlie) and Alice  use the basis $ \lambda $ for intensity $k$ .

\textit{4. Parameter estimation.}  Bob (Charlie) and Alice announce the bit information of their detection events measured under the $X$ basis. Then, they determine the number of error pulses $ m_{X, k}$, where both Bob (Charlie) and Alice, under intensity $k$, utilize the $ X $ basis and obtain inconsistent bit values. Subsequently, they use the 1-decoy-state method\cite{2018-Rusca} to estimate the lower bounds of vacuum events $s_{Z,0}^l$ and single-photon events $s_{Z,1}^l$, as well as the upper bound of the phase error rate of single-photon events  $\phi _{Z}^u$ associated with the $ Z $ basis.

\textit{5. Error correction and verification.} Bob ( Charlie) and Alice utilize the bit information measured in the Z basis to generate the raw key. They reveal $\lambda _{EC}=n_Z f h(E_{Z})$ bits of information to perform an error correction step capable of correcting errors for the expected quantum bit error rate (QBER) $ E_Z$, where $ f $ represents the error correction inefficiency function. To ensure that both parties share a pair of identical keys with a correctness of  $\varepsilon _{\rm{cor}}$, they execute an error verification step using two universal hash functions, which publish $\left \lceil \log_{2}{1/\varepsilon _{\rm{cor}}}  \right \rceil $ bits of information.

By executing steps 1-5, Bob (Charlie) and Alice distribute a pair of identical raw keys of size $ n_Z $, denoted as $ K_b $ ($ K_c $), respectively. Subsequently, Alice generates a new key string $ K_a= K_b \oplus K_c$ by performing an XOR operation on her key strings $ K_b $ and $ K_c $. If the signing of a message $ M $ requires $ 2L $ bits of the key, Alice randomly distills $ 2L $ key bits from her key string $ K_a $ to create two $ L $-bit key strings $ \left \{ \mathbb{X_\mathit{a} }, \mathbb{Y_\mathit{a} } \right \} $ . She then announces the positions of these bits to Bob and Charlie via an authenticated channel. Upon receiving this information, Bob and Charlie extract their corresponding key strings $ \left \{ \mathbb{X_\mathit{b} }, \mathbb{Y_\mathit{b} } \right \} $ and $ \left \{ \mathbb{X_\mathit{c} }, \mathbb{Y_\mathit{c} } \right \} $  from $ K_b$ and $ K_c $ at the positions indicated by Alice, ensuring that the relationships $ \mathbb{X_\mathit{a} } = \mathbb{X_\mathit{b} } \oplus \mathbb{X_\mathit{c} }$ and $\mathbb{Y_\mathit{a} } = \mathbb{Y_\mathit{b} } \oplus \mathbb{Y_\mathit{c} } $ are satisfied. The remaining portions of the key strings $ K_a $, $ K_b$, and $ K_c$ can be re-randomly extracted for subsequent signing tasks until they are fully utilized.

\begin{figure*}[h!]
	\begin{center}
		\begin{tabular}{c}
			\includegraphics[height=5.5cm]{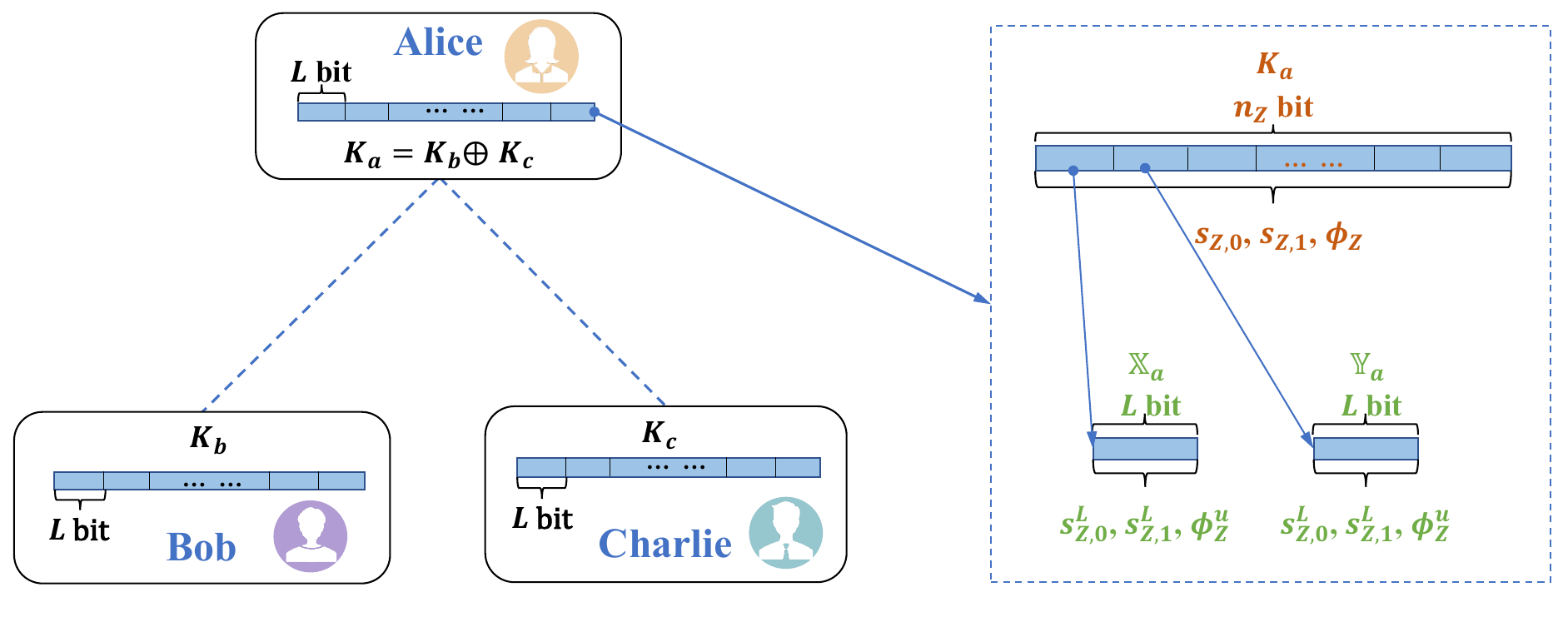}
		\end{tabular}
	\end{center}
	\caption 
	{ \label{Fin}
		An illustration of finite key analysis and corresponding parameters. In the distribution stage each of Alice, Bob, and Charlie holds an $n_Z$-bit key string that will later be divided into $L$-bit sub-strings that are used for generating the signature. The $n_Z$-bit key string contains $s_{Z,0}$ vacuum events and $s_{Z,1}$ single-photon events, and the phase error rate is $\phi_Z$. For an $L$-bit sub-string, it contains $s_{Z,0}^L$ vacuum events and $s_{Z,1}^L$ single-photon events, and the upper bound of phase error rate is  $\phi_Z^u$. }
\end{figure*}

%  To sign a message $M$, \blk Alice, Bob, and Charlie each select two $L$-bit key sequences from corresponding positions in their respective key sequences $K_a$, $K_b$ and $K_c$. These selected key sequences are denoted as $\{X_a,Y_a\}$, $\{X_b,Y_b\}$ and $\{X_c,Y_c\}$.

\vspace{0.3cm} 
\noindent 
\textbf{Messaging stage.} Upon completion of the distribution stage, Alice, Bob, and Charlie can proceed to sign the message $M $ at any time using the following steps:

\textit{1. Signing of Alice.} Alice generates an $L$-bit string  $P$ using a local quantum random number generator. Then, she constructs a random polynomial using $P$ as the seed for a hash function. Subsequently, she inputs the message to be signed (denoted as $M$) into a special group of hash functions called generalized division hash functions $H_P(x)$ to derive the digest $Dig=H_P(M)$.  Then, she encrypts the digest using the key string $ \mathbb{X_\mathit{a} }$ to obtain the signature $Sig=Dig \oplus \mathbb{X_\mathit{a} }$, while also encrypting the random bit string $P$ using $ \mathbb{Y_\mathit{a} }$ to obtain $P_a=P\oplus \mathbb{Y_\mathit{a} }$.  Subsequently, she transmits $ \{ Sig, M,  P_a \} $ through an authenticated channel to Bob. 

\textit{2. Transference.}  Bob utilizes an authenticated classical channel to transmit the received $ \{ Sig, M,  P_a \}$ along with his key bit strings $ \left \{ \mathbb{X_\mathit{b} }, \mathbb{Y_\mathit{b} } \right \} $  to Charlie. Subsequently, Charlie also employs an authenticated channel to send his key bit strings $ \left \{ \mathbb{X_\mathit{c} }, \mathbb{Y_\mathit{c} } \right \} $  to Bob.

\textit{3. Verification.} Bob and Charlie will independently verify the signature. To accomplish this, Bob creates two new key strings $\{K_{\mathbb{X_\mathit{b} }}=\mathbb{X_\mathit{b} } \oplus \mathbb{X_\mathit{c} }$, $K_{\mathbb{Y_\mathit{b} }}=\mathbb{Y_\mathit{b} } \oplus \mathbb{Y_\mathit{c} }\}$ through XOR operations. Then, he uses $K_{\mathbb{X_\mathit{b} }}$ and $K_{\mathbb{Y_\mathit{b} }}$ to decrypt the expected digest $Dig_b^{'}=Sig \oplus K_{\mathbb{X_\mathit{b} }}$ and the random sequence $P_b=P_a \oplus K_{\mathbb{Y_\mathit{b} }}$ via XOR operations, respectively. Following this, he constructs a hash function using $P_b$ and applies the hash operation to obtain the actual digest $Dig_b$ of the message $M $ . If the actual digest is equal to the expected digest, Bob will accept the signature; otherwise, he will reject the signature and abort the protocol. 

If Bob accepts the signature, Charlie will also perform similar steps for verification. She generates a new key bit string $\{K_{\mathbb{X_\mathit{c} }}=\mathbb{X_\mathit{b} } \oplus \mathbb{X_\mathit{c} }$, $K_{\mathbb{Y_\mathit{c} }}=\mathbb{Y_\mathit{b} } \oplus \mathbb{Y_\mathit{c} }\}$ using Bob's key and his own key, and uses this new key string to decrypt the expected digest $Dig_c^{'}=Sig \oplus K_{\mathbb{X_\mathit{c} }}$ and the random sequence $P_c=P_a \oplus K_{\mathbb{Y_\mathit{c} }}$ through XOR operations. Then, he constructs a hash function using $ P_c $ and applies the hash operation to obtain the actual digest $Dig_c$ of $ M $. The signature is accepted if the two digests are the same; otherwise, the signature is rejected.

\section{Security analysis of OUTH-QDS}\label{sec_ana}

Here, we will analyze the security parameters of our protocol resisting robustness, forgery, and non-repudiation. In this protocol, we use imperfect keys with privacy leakage to encrypt the hash value and hash function. Thus, any possible attackers may obtain partial information on the keys.  In the finite-size scenario, Alice, Bob, and Charlie generate the key string $K\in  \left \{ \mathbb{X_\mathit{a} }, \mathbb{Y_\mathit{a} }, \mathbb{X_\mathit{b} }, \mathbb{Y_\mathit{b} }, \mathbb{X_\mathit{c} }, \mathbb{Y_\mathit{c} }  \right \}$ using KGP. $E$ as an eavesdropper's system after performing optimal measurement strategy on $K$. According to the decoy-state method\cite{2014-Lim, 2018-Rusca}, the min-entropy of $K$ and $E$ is
\begin{equation}\label{eq1}
	\begin{split}
		H_{\rm{min}}(K|E) \leq s_{Z, 0}^{L} +s_{Z, 1}^{L}\left(1-h\left(\phi_{Z}^{L,u}\right)\right)\\
		-\frac{L}{n_Z}\left(\lambda_{\mathrm{EC}}+\log _{2}\left(2 / \epsilon_{\mathrm{cor}}\right)\right),        	
	\end{split}  
\end{equation}
where $s_{Z,0}^{L}$  is the lower bound of the vacuum state event that Alice detects Bob (Charlie) sending under the Z basis in an $L$-bit string, $s_{Z,1}^{L} $ is the lower bound of the single-photon state event that Alice detects Bob (Charlie) sending under the Z basis in an $L$-bit string, $\phi _{Z}^{L,u} $ is the upper bound of the phase error rate of the single-photon state event that Alice detects Bob (Charlie) sending under the $Z$ basis in an $L$-bit string. $\lambda  _{\rm{EC}}$ is the number of bits revealed in the error correction process.  $\epsilon_{\rm{cor}}=\varepsilon _{\rm{hash}}$ is the parameter for evaluating correctness. The $h\left ( x \right )=-x\log_{2}{x}-\left ( 1-x \right )\log_{2}{\left ( 1-x \right ) }$ is the binary Shannon entropy function. 

Furthermore, we will demonstrate how to derive the different parameters in Eq. \ref{eq1} from experimental data. To visually depict this process, we partition the parameter estimation into two segments, as shown in Fig. \ref{Fin}. In the protocol, Alice first shares an $n_Z$-bit string $K_b$ with Bob and $K_c$ with charlie, and generates $K_a=K_b\oplus K_c$ as her own key string. Three parties then disturb the order of their strings and divide them into $L$-bit sub-strings.  Parameter $s_{Z,0}^{L}$, $s_{Z,1}^{L}$, $\phi _{Z}^{L,u}$ is about an $L$-bit sub-string and can be estimated from parameters that characterize the $n_Z$-bit string.

For events $n_Z$ measured in the $Z$ basis during the distribution process, the lower bounds of vacuum events and single-photon events are estimated as follows\cite{2018-Rusca}:
\begin{equation}\label{sz0}
	\begin{split}
		s_{{Z}, 0} \geq s_{{Z}, 0}^{l}:=\frac{\tau_{0}}{\mu-\nu}\left(\mu n_{{Z}, \nu}^{-}-\nu n_{{Z}, \mu}^{+}\right),   	
	\end{split}  
\end{equation}
\begin{equation}\label{sz1}
	\begin{split} 
		s_{{Z}, 1} \geq s_{{Z}, 1}^{l}:=  \frac{\tau_{1} \mu}{\nu\left(\mu-\nu\right)}\left(n_{{Z}, \nu}^{-}-\frac{\nu^{2}}{\mu^{2}} n_{{Z}, \mu}^{+}-\frac{\left(\mu^{2}-\nu^{2}\right)}{\mu^{2}} \frac{s_{{Z}, 0}^{u}}{\tau_{0}}\right),    	
	\end{split}  
\end{equation}
where $\tau_{n}: = {\textstyle \sum_{k\in  \mathcal{K} }} p_{k}e^{-k }k^{n}/n!$ represents the total probability of sending an $n$-photon state, and $n_{Z,k}^{\pm }$ denotes the correction for the observed values in intensity $k\in \left \{ \mu, \nu \right \} $ due to statistical fluctuations considered through the Hoeffding inequality\cite{1963-Hoeffding}, given by
\begin{equation}\label{Hoe}
	\begin{split} 
		n_{Z,k}^{\pm }: = \frac{e^k}{P_k}\left ( n_{Z,k}\pm \sqrt{\frac{n_{Z}}{2}\mathrm{log} \frac{1}{\varepsilon _{SF}}  }  \right ) ,	
	\end{split}  
\end{equation}
and the upper bound of vacuum events under finite data size is given by 
\begin{equation}\label{sz0u}
	\begin{split} 
		s_{{Z}, 0}\le s_{{Z}, 0}^{u} : = 2\left ( \tau_{0}  \frac{e^k}{P_k}\left ( m_{Z,k}+ \sqrt{\frac{m_{Z}}{2}\mathrm{log} \frac{1}{\varepsilon _{SF}}  }  \right ) + \sqrt{\frac{n_{Z}}{2}\mathrm{log} \frac{1}{\varepsilon _{SF}}  }  \right ),	
	\end{split}  
\end{equation}
$ m_{Z,k}$ represents the total number of error events in intensity $k$. $ n_{Z}={\textstyle \sum_{k\in  \mathcal{K} }}n_{Z,k}$ and $ m_{Z}={\textstyle \sum_{k\in  \mathcal{K} }}m_{Z,k}$ respectively represent the total number of events and the number of error events in the $Z$ basis. $\varepsilon _{SF}$ represents the failure probability of statistic fluctuation that is set as $10^{-10}$.

\begin{table*}[t]
	
	\caption{ Experimental parameters and results on the Bob (Charlie)-Alice link. A, B, and C represent users Alice, Bob, and Charlie, respectively. $T$ and Loss represent the transmission distance and loss of the Bob (Charlie)-Alice fiber link, respectively. $\mu~(\nu)$ is the intensity of the signal (decoy) state sent by Bob and Charlie, and $P_{\mu}$ ($P_{\nu}$) represents the probability of selecting intensity $\mu~(\nu)$. $P_{Z}$ ($P_{X}$) represents the probability of Bob and Charlie selecting the $Z$ ($X$) basis. $n_{Z,\mu}$ ($n_{Z,\nu}$) represents the total number of detection events when Bob and Charlie send the signal (decoy) state in the $Z$ basis and Alice measures in the $Z$ basis. $m_{Z,\mu}$ ($m_{Z,\nu}$) represents the total number of detection errors when Bob and Charlie send the signal (decoy) state in the $Z$ basis and Alice measures in the $Z$ basis. $m_{X,\mu}$ ($m_{X,\nu}$) represents the total number of detection errors when Bob and Charlie send the signal (decoy) state in the $X$ basis and Alice measures in the $X$ basis. $n_{Z}$ represents the total accumulated detection events in the $Z$ basis after basis reconciliation, $t$ represents the time required to accumulate the data size of $n_{Z}$, $s_{Z,1}^{l}$ represents the lower bound of the single-photon detection events in the $Z$ basis, $E_{Z}$ represents the quantum bit error rate in the $Z$ basis, $\phi _{Z}^{u}$ represents the upper bound of the phase error rate in the $Z$ basis, $\lambda_{\rm EC}$ represents the number of bits leaked in the error correction step, $L$ represents the signature length, $\epsilon$ is the security parameter, and $R_S$ represents the optimal signature rate achievable at each link. Note that a QDS process requires two links and the actual signature rate is the low one of the two rates.}
	\centering
	\resizebox{\linewidth}{!}{
		
		\begin{tabular}{cccccccccccccc} 
			
			\hline\hline
			\rule[-1ex]{0pt}{3.5ex}$T$ (km) &Link& Loss (dB) &$\mu$ &$\nu$ &$P_{\mu}$ &$P_{\nu}$ & $P_{Z}$ &$P_{X}$ &$n_{Z,\mu}$& $m_{Z,\mu}$ & $n_{X,\mu}$ &$m_{X,\mu}$  &$n_{Z,\nu}$
			\\ \hline
			\rule[-1ex]{0pt}{3.5ex}\multirow{2}{*}{50}    & A-B & 9.72   & 0.601 & 0.147 & 0.807 & 0.193    & 0.947    & 0.053   & 9449854 & 51823  & 516479  & 2387&550146  
			\\
			& A-C & 9.72   & 0.481 & 0.127 & 0.775 & 0.225    & 0.935    & 0.065   & 9280871 & 130700 & 561044  & 8154  &719129
			\\ \hline
			\rule[-1ex]{0pt}{3.5ex}\multirow{2}{*}{100}    & A-B & 19.24  & 0.599 & 0.147 & 0.807 & 0.193    & 0.948    & 0.052   & 9448344 & 54494  & 511670  & 3523&551656
			\\
			& A-C & 19.24  & 0.479 & 0.127 & 0.775 & 0.225    & 0.935    & 0.065   & 9279475 & 119659 & 597173  & 5052 & 720525
			\\ \hline
			\rule[-1ex]{0pt}{3.5ex}\multirow{2}{*}{150}    & A-B & 29.16  & 0.597 & 0.146 & 0.808 & 0.192    & 0.947    & 0.053   & 9449307 & 63862  & 525201  & 1935&550693
			\\
			& A-C & 29.16  & 0.478 & 0.127 & 0.773 & 0.227    & 0.934    & 0.066   & 9279885 & 95510  & 598243  & 8926  &720115
			\\ \hline
			\rule[-1ex]{0pt}{3.5ex}\multirow{2}{*}{200}    & A-B & 39.23  & 0.593 & 0.144 & 0.798 & 0.202    & 0.935    & 0.065   & 9411383 & 92289  & 638373  & 6601&588617
			\\
			& A-C & 39.23  & 0.472 & 0.123 & 0.768 & 0.233    & 0.918    & 0.082   & 9240824 & 186323 & 810524  & 14213&759176
			\\ \hline\hline
			\\ \hline\hline

			\rule[-1ex]{0pt}{3.5ex}$T$ (km) &Link&  $m_{Z,\nu}$ & $n_{X,\nu}$ & $m_{X,\nu}$ & $n_{Z}$ & $t$ (s) & $s_{Z,1}^{l} $ & $E_{Z}$ &  $\phi _{Z}^{u}$ & $\lambda_{\rm EC}$ &$L$ &$\epsilon$& $R_S$ (tps) 
			\\ \hline
			\rule[-1ex]{0pt}{3.5ex}\multirow{2}{*}{50}    & A-B & 4781  & 27760 & 251   & $10^7$  & 74.8     & 4878658 & 0.566\% & 0.0210 & 581997 &698&4.65$\times 10^{-8}$ &90.8 
			\\
			& A-C  & 12469 & 44603 & 1069  &$10^7$ & 105.4    & 5605116 & 1.432\% & 0.0353 & 1188112 & 844 & 4.56$\times10^{-8}$ & 56.2  
			\\ \hline
			\rule[-1ex]{0pt}{3.5ex}\multirow{2}{*}{100}    & A-B & 6631  & 28700 & 324   &$10^7$ & 662.6    & 4849083 & 0.611\% & 0.0270 & 595119 & 735 & 4.78$\times10^{-8}$ & 10.3
			\\
			& A-C  & 13943 & 39847 & 536   & $10^7$  & 981.7    & 5642925 & 1.336\% & 0.0312 & 1136046 & 783 & 4.64$\times10^{-8}$ & 6.50  
			\\ \hline
			\rule[-1ex]{0pt}{3.5ex}\multirow{2}{*}{150}    & A-B & 7326  & 28411 & 224   & $10^7$  & 6686.3   & 4872943 & 0.712\% & 0.0156 & 681965 & 718 & 4.61$\times10^{-8}$ & 1.04
			\\
			& A-C  & 13063 & 44413 & 984   &$10^7$  & 9775.2   & 5546132 & 1.086\% & 0.0436 & 967145 & 812 & 4.78$\times10^{-8}$ &  0.630  
			\\ \hline
			\rule[-1ex]{0pt}{3.5ex}\multirow{2}{*}{200}    & A-B & 18264 & 40011 & 1268  & $10^7$  & 79799.4  & 4722076 & 1.106\% & 0.0262 & 965191 &  922 & 4.77$\times10^{-8}$ &$6.46\times  10^{-2} $
			\\
			& A-C  & 31614 & 72330 & 3316  & $10^7$  & 117291.2 & 5553558 & 2.179\% & 0.0256 & 1726908 & 1029 & 4.72$\times10^{-8}$ & $4.14\times  10^{-2} $
			
			\\ \hline\hline
			
			\multicolumn{13}{l}{  Note: The detection efficiency is 60\% at 200 km and 70\% at other distances. } \\ %表注	

		\end{tabular}
	}
	\label{table_rawdate}
\end{table*}

For events measured in the $X$ basis, we utilize the following formula to estimate the phase error rate of the single-photon events in the $Z$ basis
\begin{equation}\label{phi_z}
	\begin{split} 
		\phi _{Z}: = \frac{c_{Z,1}}{s_{Z,1}} \le \frac{v_{X,1}}{s_{X,1}} +\gamma^{U} \left (s_{Z,1}, s_{X,1}, \varepsilon _{SF},\frac{v_{X,1}}{s_{X,1}} \right ),	
	\end{split}  
\end{equation}
where 
\begin{equation}
	\gamma^U( n, k, \epsilon, \lambda )=\frac{\frac{(1-2\lambda)AG}{n+k}+\sqrt{\frac{A^2G^2}{(n+k)^2}+4\lambda(1-\lambda)G}}{2+2\frac{A^2G}{(n+k)^2}},
\end{equation}
in which 
\begin{equation}
	A=\max\{n,k\},
\end{equation}
and
\begin{equation}
	G=\frac{n+k}{nk}\ln(\frac{n+k}{2\pi nk\lambda(1-\lambda)\epsilon^2}).
\end{equation}

The upper bound on the number of error bits for single-photon events in the $X$ basis can be estimated using the following formula
\begin{equation}\label{vx1}
	\begin{split} 
		v_{X,1}\le v_{X,1}^{u}=\frac{\tau _{1}}{\mu-\nu } ( m_{X,\mu}^{+}- m_{X,\nu}^{-}),	
	\end{split}  
\end{equation}
$m_{X,k}^{\pm}$ represents the observed number of errors bits for intensity  $k$ considering a scenario with finite key, and it can be expressed as
\begin{equation}\label{mxk}
	\begin{split} 
		m_{X,k}^{\pm }: = \frac{e^k}{P_k}\left ( m_{X,k}\pm \sqrt{\frac{m_{X}}{2}\mathrm{log} \frac{1}{\varepsilon _{SF}}  }  \right ) .	
	\end{split}  
\end{equation}

We can now compute the upper bound on the phase error rate of single-photon events in the $Z$ basis using the following formula
\begin{equation}\label{phi_ZU}
	\begin{split} 
		\phi _{Z}\le	\phi _{Z}^{u} : =   \frac{v_{X,1}^{u}}{s_{X,1}^{l}} +\gamma^{U} \left (s_{Z,1}^{l},s_{X,1}^{l}, \varepsilon _{SF},\frac{v_{X,1}^{u}}{s_{X,1}^{l}} \right ).	
	\end{split}  
\end{equation}
the estimation method of the lower bound $s_{X,1}^{l}$ for single-photon events in the $X$ basis is similar to the analysis of $s_{Z,1}^{l}$ in the $Z$ basis.

In the protocol Alice, Bob, and Charlie all randomly divide their keys into $L$-bit strings. This is a random sampling without replacement process, and thus the statistical fluctuation can be bounded by $\gamma ^{U}$.
Then we can obtain the bound of $s_{{Z}, 0}^L$, $s_{{Z}, 1}^L$, and $\phi _{Z}^u$ in Eq.\ref{eq1}

\begin{equation}
	\begin{aligned}
		s_{Z,0}^L\geq&L\left[{s}_{Z,0}/n_Z-	\gamma^U (L,n_Z-L, \varepsilon_{SF},{s}_{Z,0}/n_Z)\right],\\
	\end{aligned}
\end{equation}
\begin{equation}
	\begin{aligned}
		s_{Z,1}^L\geq&L\left[{s}_{Z,1}/n_Z-	\gamma^U (L, n_Z-L, \varepsilon_{SF},{s}_{Z,1}/n_Z)\right],\\
	\end{aligned}
\end{equation}
\begin{equation}
	\begin{aligned}
		\phi_{Z}^{L,u}\leq& \phi_{Z}^{u}+	\gamma^U \left({s}_{Z,1}^L,s_{Z,1}^l-{s}_{Z,1}^L, \varepsilon_{SF},\phi_{Z,1}\right).\\
	\end{aligned}
\end{equation}

The amount of information the attacker obtain can be estimated by Eq.\ref{eq1}, and the maximum probability that  an attacker successfully guesses one of the strings ${X_b,Y_b,X_c,Y_c}$ is no more than
\begin{equation}
	Pr=2^{-H_n},
\end{equation}
where $H_n=H_{\rm{min}}(K|E)$ in Eq.~\ref{eq1}, represents the unknown information to the eavesdropper within the $n$-bit string generated in the distribution stage.

Additionally, there are three security parameters in the QDS process that require analysis.

\noindent\textbf{Robustness} The honest run aborting  will occur only when Alice and Bob (or Charlie) share different key bits after the distribution stage. In our protocol Alice and Bob (Charlie) will perform error correction in distribution stage. Thus, they will share the identical final key, unless the error correction fails. The robustness bound is $\epsilon_{\rm{rob}}=2\epsilon_{\rm{cor}}$.  
%The robustness bound $\epsilon_{rob}=2\epsilon_{QKD}$, where $\epsilon_{QKD}$ is the failure probability of TP-TFQKD. In the the asymptotic case $\epsilon_{QKD}=0$, so $\epsilon_{rob}=0$

\noindent\textbf{Repudiation}
For Alice's repudiation attacks, Bob and Charlie are both honest and symmetric and they have the same new key strings. They will make the same decision for the same document and signature. In other words, when Bob rejects (accepts) the document, Charlie also rejects (accepts) it. Therefore, our QDS protocol is immune to repudiation naturally, i.e., the repudiation bound $\epsilon_{\rm{rep}}=0$.

\noindent\textbf{Forgery}
Bob forges successfully when Charlie accepts the forged document forwarded by Bob. According to our protocol, Charlie accepts the message if and only if Charlie gets the same result through one-time pad encryption and one-time hash functions. Since the signature is encrypted by one-time pad and each hash function is utilized only one time, in a specific round Bob can obtain no information about the hash function or the hash value from previous rounds.
Therefore, the probability of a successful forgery is identical to the failure probability of authentication scheme based on hashing, i.e., one finds two distinct documents to have the identical hash value~\cite{2023-Li}. 
Due to the construction of hash functions, Bob's optimal strategy is to guess the monic irreducible polynomial $p(x)$ which can be conclude if successfully guessing $X_c$. Also, since he knows that $p(x)$ is irreducible, he can guess only from all irreducible polynomials (no less than $2^{(n+2)}/n$) rather than all polynomials ($2^n$). The probability that Bob successfully guesses $p(x)$ is $n/4 Pr$.
To finish a foegery attack, Bob can generate a polynomial $g(x)$ of order no more than $m$, and construct $M' (x)=M(x)+g(x),Sig'=Sig$. As long as $p(x)$ is a factor of $g(x)$, Charlie will pass ${M',Sig'}$. Thus, Bob can guess $p(x)$ for no more than $m/8$ times.
The forgery bound is $\epsilon_{\rm for}= \frac{m}{8\cdot 2^{H_n-1}}$.

The security bound of the QDS protocol, i.e., the maximum probability that an attack is successfully performed, is defined as $\epsilon=\max\{\epsilon_{\rm{rob}},\epsilon_{\rm{rep}},\epsilon_{\rm{for}}\}$.

It can be seen that $\epsilon_{\rm{rob}}$ and $\epsilon_{\rm{rep}}$ are fixed number, and only $\epsilon_{\rm{for}}$ is decided by $H_n$ influenced by parameters such as channel loss and signature length $L$.
Thus, the variation of $\epsilon$ is mainly caused from $\epsilon_{\rm{for}}$. In other words, the selection of $L$ is constrained by ensuring $\epsilon_{\rm{for}}$ being small enough.

% \section{Security analysis of OUTH-QDS}\label{sec_ana}

\section{Detailed experimental results}\label{Exp-R}
Table~\Ref{table_rawdate} shows the detailed experimental results.

\vspace{0.5cm}
\noindent 	
\textbf{Data Availability.}~~The data that support the findings of this study are available from the corresponding author on reasonable request.

%\bibliography{Chip-QDS}
%\bibliographystyle{naturemag}

\vspace{0.5cm}
\noindent 
\textbf{Acknowledgments}~~We thank Shizhuo Li for drawing the diagram of the chip; We also thank Chenyu Xu for collecting experimental data.

\vspace{0.5cm}
\noindent 
\textbf{Funding}~~This study was supported by the National Natural Science Foundation of China (Nos. 12274223, 62171144,  62031024, and 62171485), the Guangxi Science Foundation (No.
2021GXNSFAA220011), the Open Fund of IPOC (BUPT) (No. IPOC2021A02)  and Innovation Project of Guangxi Graduate Education  (No. YCBZ2024002).

%\section*{Additional information}\label{MISCELLANEA}
%\vspace{-0.3cm}
%\noindent 
%\textbf{Supplementary materials}.

\vspace{0.5cm}

\vspace{0.5cm}
\noindent 
\textbf{Declaration of Competing Interest}~~The authors declare no competing interests.

\end{document}